\documentclass[a4paper,11pt]{article}

\usepackage{amsfonts,amsthm,amssymb}
\usepackage{mathtools}
\usepackage{graphicx}
\usepackage[section]{placeins}
\usepackage{subcaption}
\usepackage{caption}
\usepackage{url}
\usepackage{mathtools}
\usepackage{todonotes}
\usepackage{soul}

\newcommand{\ket}[1]{| #1 \rangle}
\newcommand{\bra}[1]{\langle #1 |}
\newcommand{\braket}[2]{\langle #1 | #2 \rangle}

\def\red#1{\textcolor{red}{#1}}

\newcommand{\neswarrow}{\mathrel{\text{\ooalign{$\swarrow$\cr$\nearrow$}}}}
\newcommand{\nwsearrow}{\mathrel{\text{\ooalign{$\nwarrow$\cr$\searrow$}}}}

\def\UA{\uparrow}
\def\DA{\downarrow}
\def\LA{\leftarrow}
\def\RA{\rightarrow}
\def\SA{\circlearrowleft}
\def\ULA{\nwarrow}
\def\DLA{\swarrow}
\def\URA{\nearrow}
\def\DRA{\searrow}
\def\LRA{\leftrightarrow}
\def\DRULA{\neswarrow}
\def\DLURA{\nwsearrow}

\newcommand{\comment}[1]{}

\title{Lackadaisical quantum walks on 2D grids with multiple marked vertices}

\author{Nikolajs Nahimovs and Raqueline A. M. Santos}
\date{\small{Center for Quantum Computer Science, Faculty of Computing, University of Latvia} \\ 
\small{Raina bulv. 19, Riga, LV-1586, Latvia}\\
\small{\texttt{nikolajs.nahimovs@lu.lv, rsantos@lu.lv}}}

\begin{document}

\maketitle


\begin{abstract}

\noindent

Lackadaisical quantum walk (LQW) is a quantum analog of a classical lazy walk, where each vertex has a self-loop of weight $l$.
For a regular $\sqrt{N}\times\sqrt{N}$ 2D grid LQW can find a single marked vertex with $O(1)$ probability in $O(\sqrt{N\log N})$ steps using $l = d/N$, where $d$ is the degree of the vertices of the grid \cite{Hoyer:2020}.
For multiple marked vertices, however, $l = d/N$ is not optimal as the success probability decreases with the increase of the number of marked vertices \cite{Nahimovs:2019}.
In this paper, we numerically study search by LQW for different types of 2D grids -- triangular, rectangular and honeycomb -- with multiple marked vertices. We show that in all cases the weight $l = m\cdot d/N$, where $m$ is the number of marked vertices, still leads to $O(1)$ success probability.

\end{abstract}

 
\section{Introduction}

Quantum walks are quantum counterparts of classical random walks \cite{Portugal:2013}. 
Similarly to classical random walks, there are two types of quantum walks: discrete-time quantum walks (DTQW),  introduced by Aharonov~{\it et al.}~\cite{Aharonov:1993}, and continuous-time quantum walks (CTQW), introduced by Farhi~{\it et al.}~\cite{Farhi:1998}.
For the discrete-time version, the step of the quantum walk is usually given by two operators -- coin and shift -- which are applied repeatedly. 
The coin operator acts on the internal state of the walker and rearranges the amplitudes of going to adjacent vertices. The shift operator moves the walker between the adjacent vertices.

Quantum walks have been useful for designing algorithms for a variety of search problems~\cite{Nagaj:2011}.
To solve a search problem using quantum walks, we introduce the notion of marked elements (vertices), corresponding to elements of the search space that we want to find.
We perform a quantum walk on the search space with one transition rule at the unmarked vertices, and another transition rule at the marked vertices. If this process is set up properly, it leads to a quantum state in which the marked vertices have higher probability than the unmarked ones. This method of search using quantum walks was first introduced in \cite{Shenvi:2003} and has been used many times since then.

The concept of lackadaisical quantum walk (LQW), i.e. quantum walk with self-loops, was first studied for DTQW on the one-dimensional line \cite{Norio:2005,Stefanak:2014}. Later it was successfully applied to improve the DTQW based search on the complete graph~\cite{Wong:2015a} and two-dimensional rectangular grid~\cite{Wong:2018}. For the rectangular 2D grid LQW gives $O(\sqrt{\log{N}})$ speed-up over the non-lackadaisical algorithm of \cite{Ambainis:2005}\footnote{There are also other methods to achieve similar speed-up, e.g. controlling the quantum walk using an ancilla qubit \cite{Tulsi:2008} or classically searching the neighbourhood of the found vertex~\cite{Nahimovs:2013}.}.

The running time of the lackadaisical walk heavily depends on the weight of the self-loop. For a regular 2D grid with a single marked vertex, the optimal weight of the self-loop is $l = d/N$, where $d$ is the degree of the vertices of the grid \cite{Hoyer:2020}.
For multiple marked vertices, however, $l = d/N$ is not optimal as the success probability decreases with the increase of the number of marked vertices \cite{Nahimovs:2019}. 

There are a few papers studying LQW search on the rectangular 2D grid with multiple marked vertices.
Saha {\it et al.}~\cite{Saha:2018} showed that if $m$ marked vertices are arranged in a $\sqrt{m}\times\sqrt{m}$ cluster, one should use $l=\frac{4}{N\left(m + \lfloor{\frac{\sqrt{m}}{2}}\rfloor\right)}$.
Giri and Korepin~\cite{Giri:2020} numerically found suitable choices of $l$ for which it is possible to find up to 6 marked vertices in $O\left(\sqrt{\frac{N}{m}\log \frac{N}{m}}\right)$
time steps with $O(1)$ success probability. Nahimovs~\cite{Nahimovs:2019} have demonstrated the existence of exceptional configurations
of marked vertices (i.e. configurations of marked vertices for which the probability of finding a marked vertex does not grow with the number of steps) and proposed two values of $l$ of the form $\frac{4(m-O(m))}{N}$ which results in $O(1)$ success probability.
Carvalho \emph{et al.}~\cite{Carvalho:2020} numerically showed that for $N\gg m$ the success probability is inversely proportional to the density of marked vertices and directly proportional to the relative distance between the marked vertices.

In this paper, we numerically study search by LQW for different types of 2D grids -- triangular, rectangular and honeycomb -- with multiple marked vertices. We show that in all cases the weight $l = m\cdot d/N$, where $m$ is the number of marked vertices, leads to $O(1)$ success probability.
The results are obtained from numerical simulations. The .NET code used to simulate the quantum walk search algorithms is available on GitHub \cite{Simulator}.

The paper is organized as follows. In Section \ref{sec:definitions}, we define the lackadaisical quantum walk and how we can do search on rectangular, triangular and honeycomb 2D grids. In Section~\ref{sec:analysis}, we find a suitable option for the self-loop weight and analyse its behavior when searching an arbitrary set of marked vertices. And we draw our conclusions in Section~\ref{sec:conclusions}.


\section{Lackadaisical Quantum walks on two-dimensional grids}\label{sec:definitions}

Consider a two-dimensional $\{$triangular, rectangular, honeycomb$\}$ grid of size $\sqrt{N}\times\sqrt{N}$ with periodic (torus-like) boundary conditions.  
The vertices of the grid are labeled by the coordinates $(x,y)$ for $x, y \in \{0,\dots,\sqrt{N}-1\}$.
The coordinates define a set of state vectors, $\{\ket{x,y}: x, y \in \{0,\dots,\sqrt{N}-1\}\}$, which span the $N$-dimensional Hilbert space ${\cal{H_P}}$ associated with the position. 
Let $d$ be the degree of the grid.
The lackadaisical quantum walk~\cite{Wong:2015} has an additional self-loop of weight $l$ in each vertex. Then, the Hilbert space associated with the directions the walker can face is a $d+1$-dimensional Hilbert space. 
We refer to it as the coin subspace ${\cal{H}_C}$. Therefore, the Hilbert space of the lackadaisical quantum walk is $\mathcal{H} = {\cal{H_P}}\otimes{\cal{H_C}} = \mathbb{C}^N\otimes \mathbb{C}^{d+1}$. 

The evolution of a state of the walk (without searching) is driven by the unitary operator $U = S\cdot (I_N \otimes C)$, where $S$ is the flip-flop shift operator~\cite{Ambainis:2005}
and $C$ is the coin operator, given by
\begin{equation}
C = 2 \ket{s_c}\bra{s_c} - I_{d+1},
\end{equation}
where 
$$
\ket{s_c} = \frac{1}{\sqrt{d+l}}\sum_{i=0}^{d-1}\ket{i}+\sqrt{l}\ket{\SA}.
$$
Notice that when $l=0$ we obtain the regular (non-lackadaisical) quantum walk.
Following we draw up the specifics for the lackadaisical quantum walk for each type of grid.


\subsection{The rectangular grid}
\begin{figure}[!htb]
\centering
\includegraphics[scale=0.8]{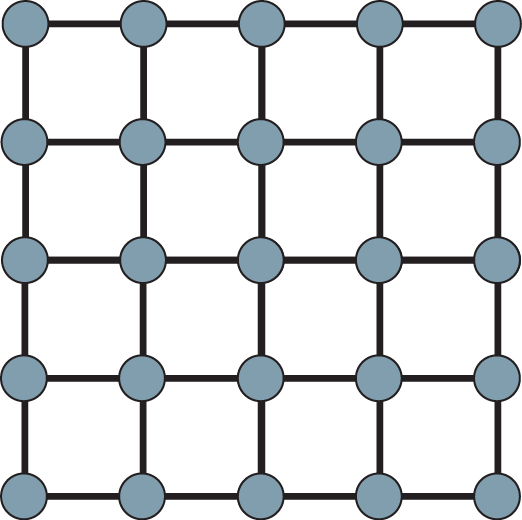}
\caption{The two-dimensional rectangular grid.}
\label{fig:rect}
\end{figure}

The coin subspace of the walk (see Fig.~\ref{fig:rect}) is a 5-dimensional Hilbert space spanned by the set of states $\{\ket{c}: c\in \{\UA,\DA,\LA,\RA,\SA \}\}$. 
The shift operator acts as 
\begin{eqnarray*}
S\ket{x,y,\UA} & = & \ket{x,y+1,\DA}, 
\\
S\ket{x,y,\DA} & = & \ket{x,y-1,\UA}, 
\\
S\ket{x,y,\LA} & = & \ket{x-1,y,\RA}, 
\\
S\ket{x,y,\RA} & = & \ket{x+1,y,\LA}, 
\\
S\ket{x,y,\SA} &=& \ket{x,y,\SA}.
\end{eqnarray*}

\subsection{The triangular grid}
The coin subspace of the walk is a 7-dimensional Hilbert space spanned by the set of states $\{\ket{c}: c\in \{\ULA,\URA,\LA,\RA,\DLA,\DRA,\SA\}\}$.
\begin{figure}[!htb]
\centering
\subcaptionbox{}{\includegraphics[scale=0.8]{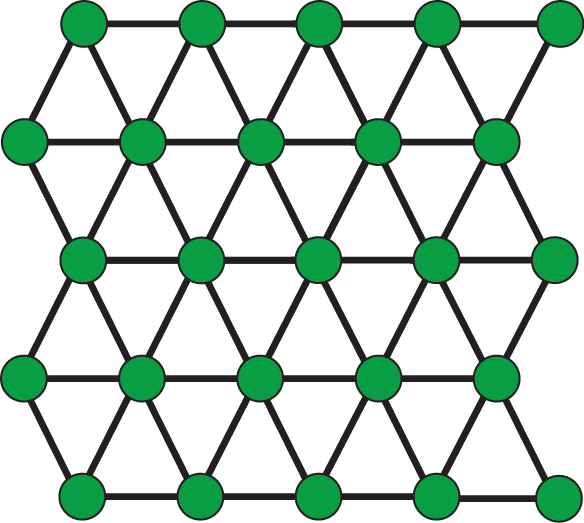}}\hspace{2cm}
\subcaptionbox{}{\includegraphics[scale=0.8]{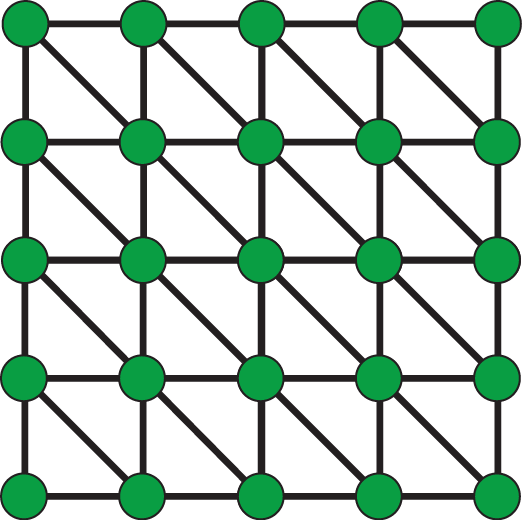}}\\
\subcaptionbox{}{\includegraphics[scale=1]{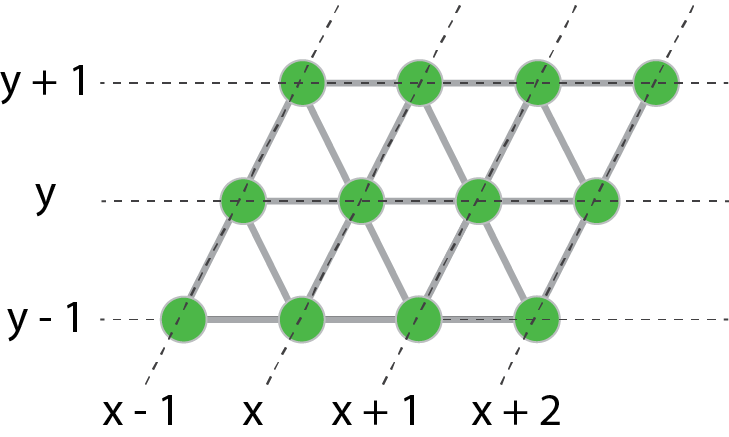}}
\caption{(a) The two-dimensional triangular grid and (b)-(c) its mapping to the rectangular grid.}
\label{fig:tri_on_rect}
\end{figure}

There exists a simple mapping from the triangular to rectangular grid as shown on Fig.~\ref{fig:tri_on_rect}.
This allows us to label vertices of the grid by the coordinates $(x,y)$ for $x, y \in \{0,\dots,\sqrt{N}-1\}$.
The shift operator $S$ acts as
\begin{eqnarray*}
S\ket{x,y,\ULA} & = & \ket{x-1,y+1,\DRA}, \\
S\ket{x,y,\DRA} & = & \ket{x+1,y-1,\ULA}, \nonumber \\
S\ket{x,y,\LA} & = & \ket{x-1,y,\RA}, \nonumber \\
S\ket{x,y,\RA} & = & \ket{x+1,y,\LA}, \nonumber \\
S\ket{x,y,\DLA} & = & \ket{x,y-1,\URA}, \nonumber \\
S\ket{x,y,\URA} & = & \ket{x,y+1,\DLA}, \nonumber \\
S\ket{x,y,\SA} &=& \ket{x,y,\SA}.
\end{eqnarray*}

\subsection{The honeycomb grid}
For the honeycomb grid there are two types of vertices, having either $\{\RA,\ULA,\DLA,\SA\}$ or $\{\LA,\DRA,\URA,\SA\}$ directions. 
We can use a $4$-dimensional coin space $\{\LRA,\DRULA,\DLURA,\SA\}$ which corresponds to either $\{\RA,\ULA,\DLA,\SA\}$ or $\{\LA,\DRA,\URA,\SA\}$ depending on the type of vertex.

\begin{figure}[!htb]
\centering
\subcaptionbox{}{\includegraphics[scale=0.8]{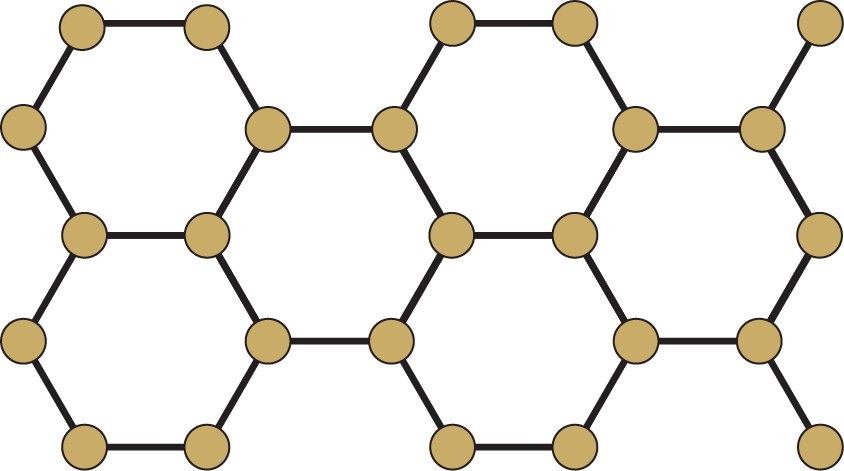}}\hspace{2cm}
\subcaptionbox{}{\includegraphics[scale=0.8]{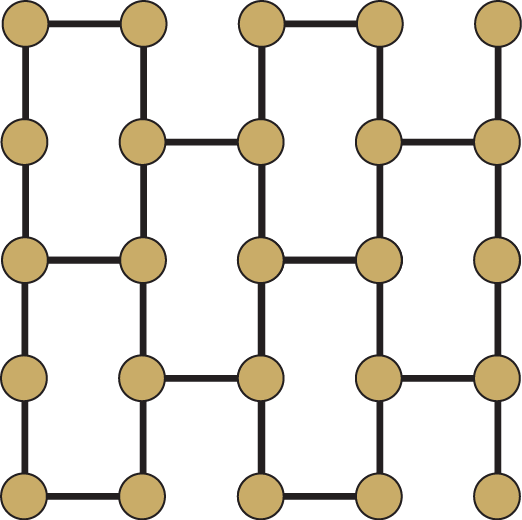}}
\subcaptionbox{}{\includegraphics[scale=1]{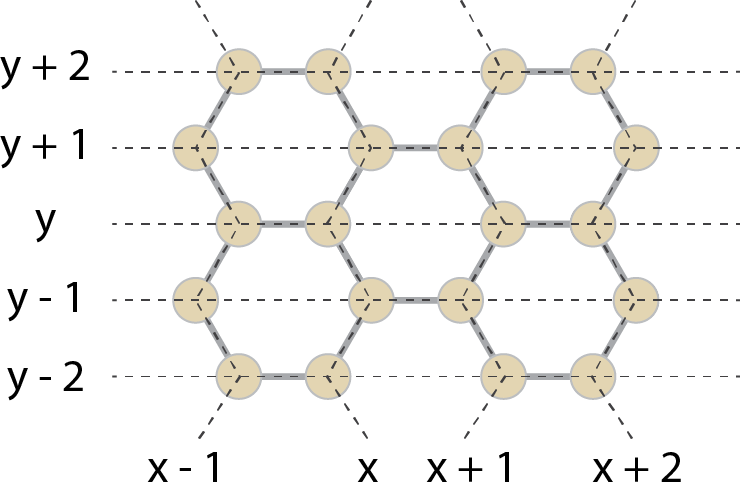}}
\caption{(a) The two-dimensional honeycomb grid and (b)-(c) its mapping to the rectangular grid.}
\label{fig:hex_on_rect}
\end{figure}

There exists a simple mapping from the honeycomb to the rectangular grid as shown on Fig.~\ref{fig:hex_on_rect}. 
This allows us to label the vertices of the grid by the coordinates $(x,y)$ for $x, y \in \{0,\dots,\sqrt{N}-1\}$.
The shift operator $S$ acts on the first type of vertices as
\begin{eqnarray*}
S\ket{x,y,\ULA} & = & \ket{x,y+1,\DRA}, \\
S\ket{x,y,\DRA} & = & \ket{x,y-1,\ULA}, \nonumber \\
S\ket{x,y,\LA} & = & \ket{x-1,y,\RA}, \nonumber \\
S\ket{x,y,\SA} &=& \ket{x,y,\SA}
\end{eqnarray*}
and on the second type of vertices as
\begin{eqnarray*}
S\ket{x,y,\RA} & = & \ket{x+1,y,\LA}, \nonumber \\
S\ket{x,y,\DLA} & = & \ket{x,y-1,\URA}, \nonumber \\
S\ket{x,y,\URA} & = & \ket{x,y+1,\DLA}, \nonumber \\
S\ket{x,y,\SA} &=& \ket{x,y,\SA}.
\end{eqnarray*}

\subsection{Searching}
In order to search for marked vertices we extend the step of the algorithm, making it
$$
U' = U \cdot (Q \otimes I_{d+1}) ,
$$
where $Q$ is the query transformation which flips the sign at marked vertices, irrespective of the coin state. 
The system starts in 
\begin{equation}
\ket{\psi(0)} = \frac{1}{\sqrt{N}} \sum_{x,y=0}^{\sqrt{N}-1} \ket{x,y} \otimes \ket{s_c} ,
\end{equation}
which is the uniform superposition over vertices (but not directions). Note that $\ket{\psi(0)}$ is a 1-eigenvector of $U$ but not of $U'$.
The state of the system after $t$ steps is $\ket{\psi(t)} = U'^t \ket{\psi(0)}$. The probability of finding a marked vertex at time $t$ is given by
\begin{equation}
    p_M(t) = \bra{\psi(t)}\left(\sum_{(x,y)\in M}\ket{x,y}\bra{x,y}\otimes I_{d+1}\right)\ket{\psi(t)},
\end{equation}
where $M$ is the set of marked vertices.
If there are marked vertices, the state of the algorithm starts to deviate from $\ket{\psi(0)}$.
For the regular quantum walk ($l=0$), 
in the case of a single marked vertex, after $O(\sqrt{N\log{N}})$ steps the inner product $\braket{\psi(t)}{\psi(0)}$ becomes close to $0$.
If the state is measured at this moment, the probability of finding a marked vertex is $O(1 / \log{N})$~\cite{Ambainis:2005,Abal:2010,Abal:2012}.
By using amplitude amplification~\cite{Brassard:2002} we obtain the total running time of $O(\sqrt{N} \log{N})$ steps. For the lackadaisical quantum walk, we can find a single marked vertex with $O(1)$ probability in $O(\sqrt{N\log N})$ steps using $l = d/N$~\cite{Hoyer:2020}.


\section{Finding a suitable weight for multiple marked vertices}\label{sec:analysis}

The running time of the search by LQW depends on the weight of the self-loop $l$. For a regular 2D grid with a single marked vertex, the optimal weight of the self-loop is $l = d/N$ \cite{Hoyer:2020}. 
For multiple marked vertices, $l = d/N$ is not optimal.

In this section we analyse the effect of different values of $l$ for the three types of 2D grids: triangular, rectangular and honeycomb.
To start, let us plot the probability of finding a marked vertex for different values of the self-loop weight $l$.
\begin{figure}[!htb]
\centering
\subcaptionbox{}{\includegraphics[width=.8\textwidth]{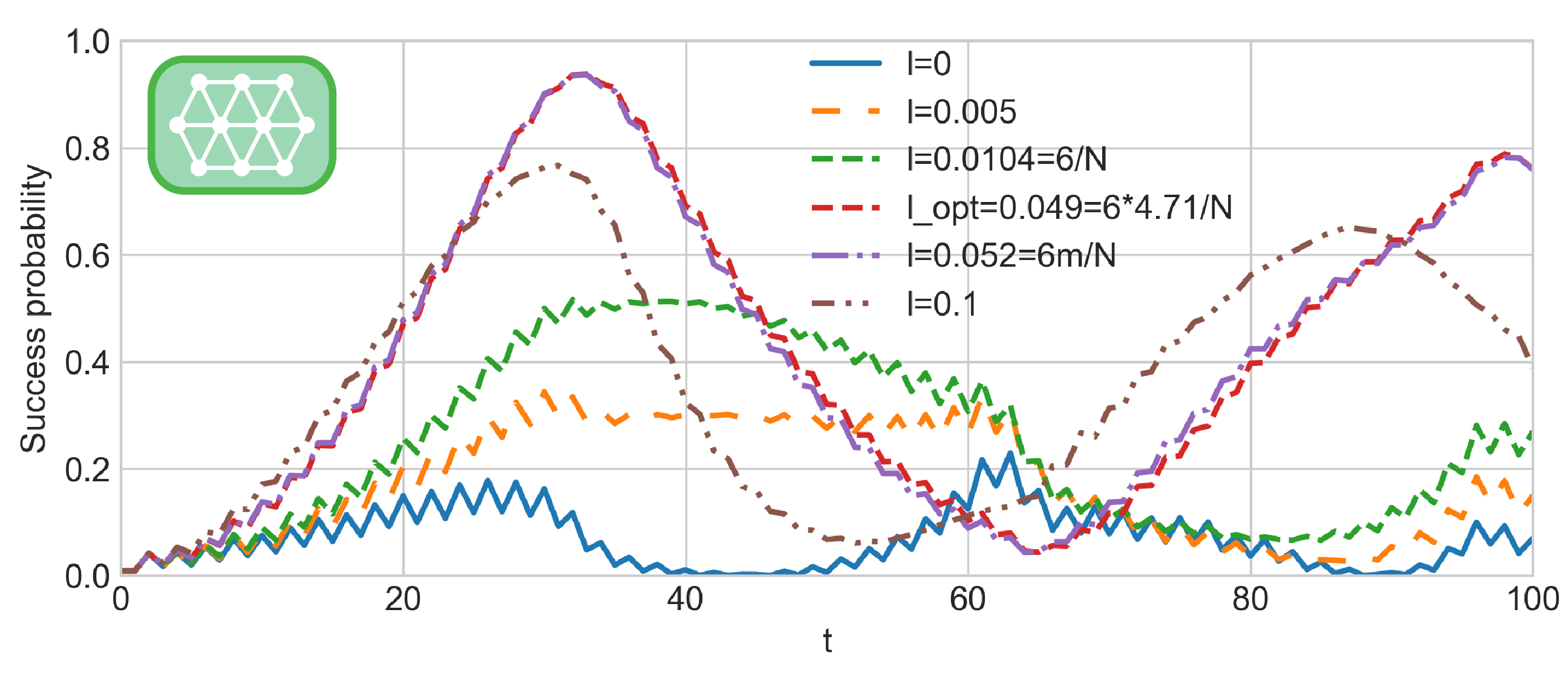}}
\hfill
\subcaptionbox{}{\includegraphics[width=.8\textwidth]{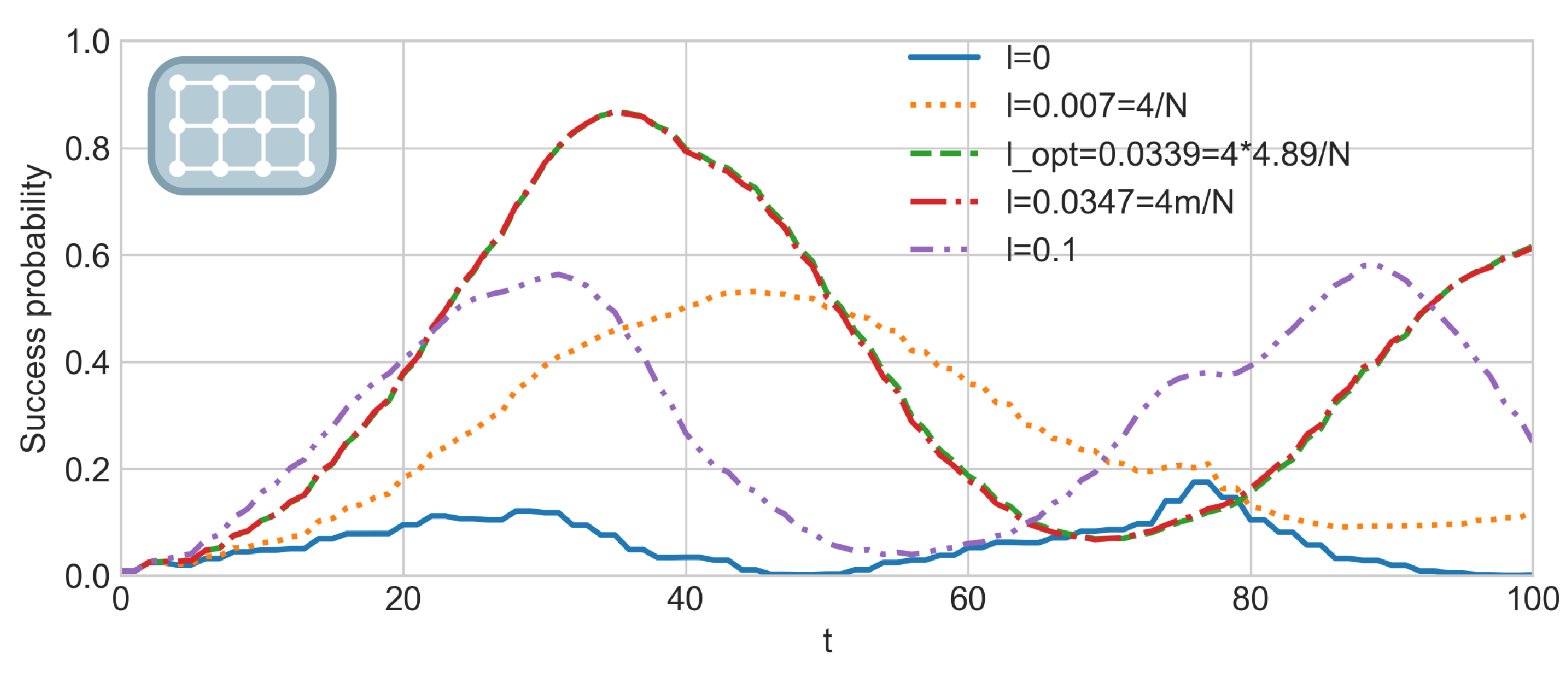}}
\hfill
\subcaptionbox{}{\includegraphics[width=.8\textwidth]{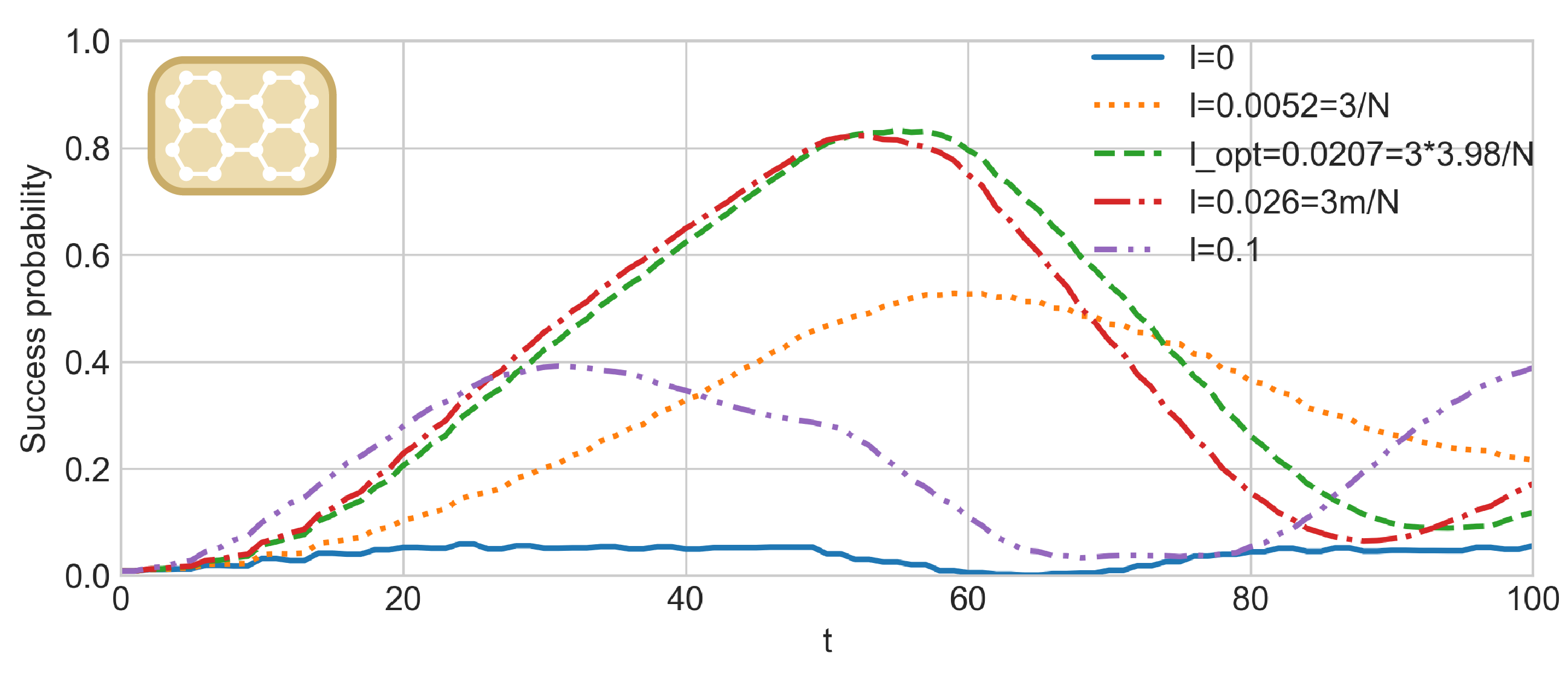}}
\caption{The probability of finding a marked vertex for different values of $l$ for the (a) triangular, (b) rectangular and (c) honeycomb grids of size $24\times 24$ with set of marked vertices $\{(0, 0),(0, 2),(0, 4),(0, 6),(0, 8)\}$. The solid blue curve is $l=0$; the dotted yellow curve is $l= d/N$; the dashed green curve is the optimal value for $l$; the dot-dashed red curve is $l=d\cdot m/N$; and the dot-dot-dashed purple curve is $l=0.1$.}
\label{fig:compare_ls}
\end{figure}
Figure~\ref{fig:compare_ls} shows the triangular (\ref{fig:compare_ls}a), rectangular (\ref{fig:compare_ls}b) and honeycomb (\ref{fig:compare_ls}c) grids of size $N = 24\times 24$ with $m=5$ marked vertices $\{(0, 0),(0, 2),(0, 4),(0, 6),(0, 8)\}$. We can observe the same behavior for all grids.
At $t=0$, the success probability is equal to $m/N = 5/576\approx 0.0087$. 
The solid blue curve $l=0$ depicts the evolution of the regular (non-lackadaisical) quantum walk. If we increase the value of $l$ the success probability starts to grow. The dotted yellow curve depicts $l=d/N$ -- the optimal value for a single marked vertex. Since we have more than one marked vertex this value of $l$ is not optimal. The green dashed curve shows the success probability, corresponding to  $l=l_{opt}\approx 0.0490, 0.0339, 0.0207$ for the triangular, rectangular and honeycomb grids, respectively.  
The dot-dashed red curve $l=d\cdot m/N$ is very close to the optimal curve, specially for the rectangular grid. By further increasing the value of $l$, the success probability starts to drop, as depicted by the purple dot-dot-dashed curve $l=0.1$. 

\begin{figure}[!htb]
\centering
\subcaptionbox{}{\includegraphics[width=0.48\textwidth]{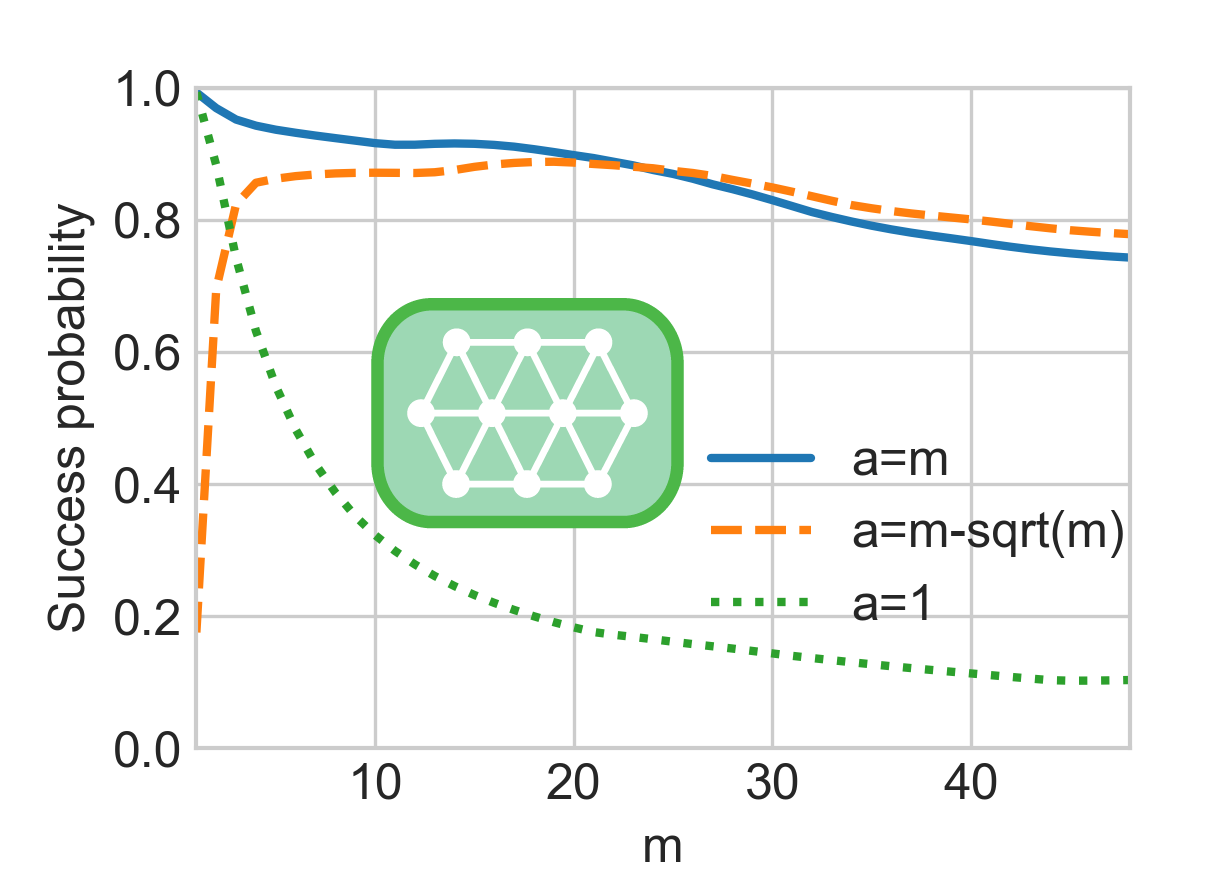}}
\hfill
\subcaptionbox{}{\includegraphics[width=0.48\textwidth]{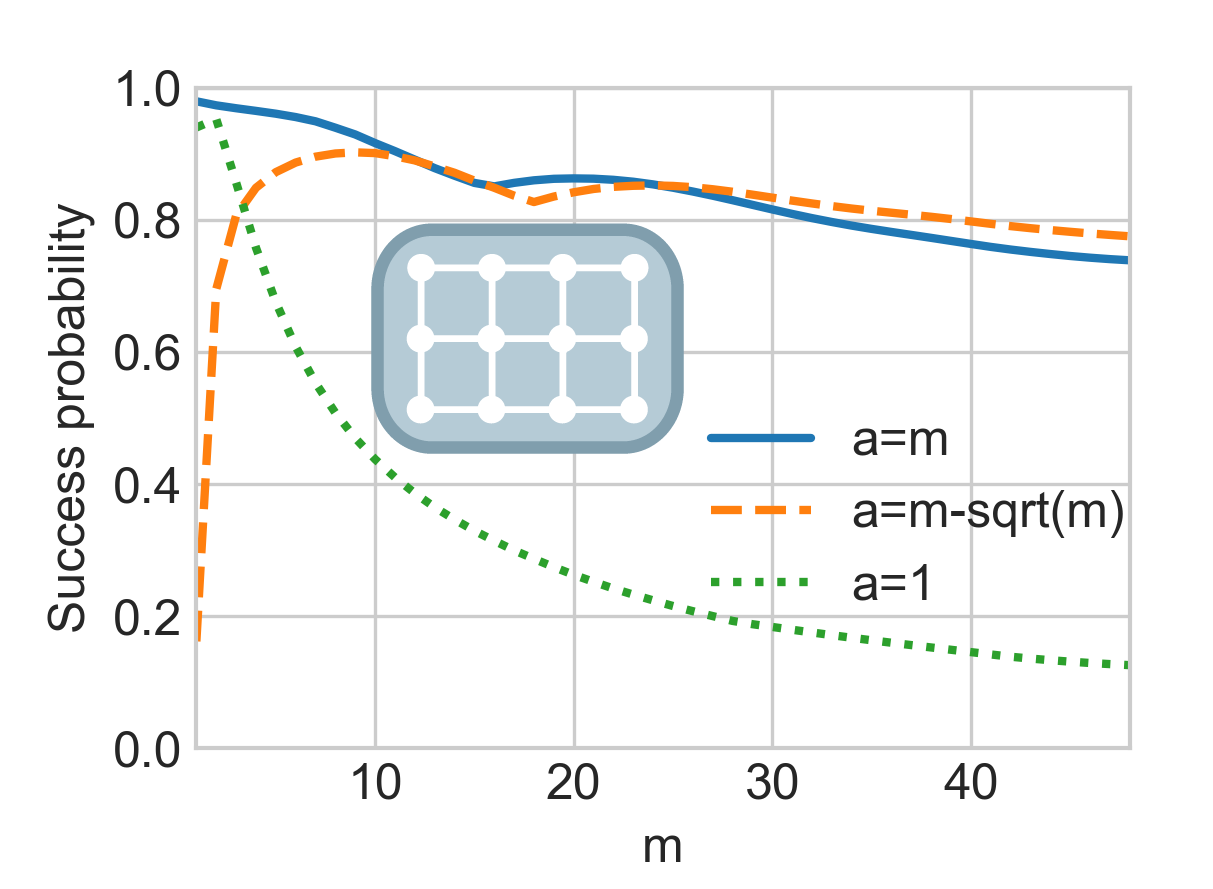}}
\hfill
\subcaptionbox{}{\includegraphics[width=0.48\textwidth]{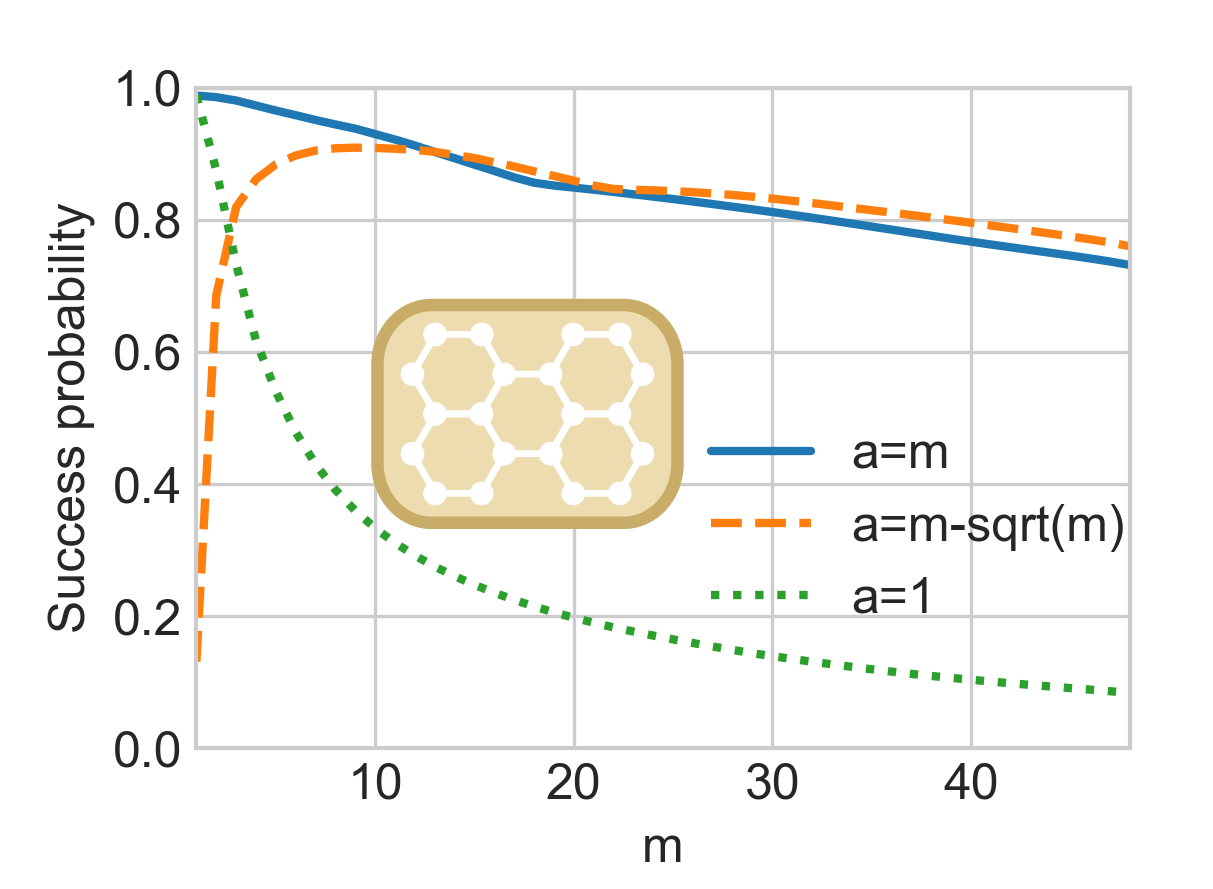}}
\caption{Comparison of the self-loop weights $l= a\cdot d/N$, where $a \in\{ m, m-\sqrt{m}, 1\}$, for $100 \times 100$ grid with set of marked vertices $M_m = \{(0, 2i)| i \in [0, m-1]\}$ for $m\in[1,48]$. The maximum success probability is shown for the (a) triangular, (b) rectangular and (c) honeycomb grids. The solid blue curves depicts $a=m$; the dashed yellow curves, $a = m-\sqrt{m}$; and the dotted green curves, $a=1$. }
\label{fig:compare_ls2}
\end{figure}

Figure~\ref{fig:compare_ls2} compares three values for the self-loop weight $l= a\cdot d/N$, $a \in\{m, m-\sqrt{m}, 1\}$. Fig.~\ref{fig:compare_ls2}a, \ref{fig:compare_ls2}b and \ref{fig:compare_ls2}c plot the maximum success probability (first peak) for the triangular, rectangular and honeycomb grids, respectively. The grids size is $N=100\times100$. The set of marked vertices is $M_m = \{(0, 2i)| i \in [0, m-1]\}$ for $m\in[1,48]$.
The dotted green curves for $a=1$ indicates a rapid decrease of the success probability as $m$ increases. It confirms that $l=d/N$ is not a good choice for improving the search algorithm. The other two values, $a=m$ and $a=m-\sqrt{m}$, look more promising. They  were first proposed in~\cite{Nahimovs:2019} for the rectangular grid. We can observe that the solid blue curves for $a=m$ and dashed yellow curves for $a=m-\sqrt{m}$ get close to each other as $m$ increases. Comparing both values, the success probability for $a=m-\sqrt{m}$ is considerably smaller for small values of $m$ and it gets slightly bigger for big values of $m$. 
\comment{\red{The plots for the total running time of the algorithm, given by Fig.~\ref{fig:compare_ls2}d, \ref{fig:compare_ls2}e and \ref{fig:compare_ls2}f for the triangular, rectangular and honeycomb grids, respectively, confirms that $l=d/N$ is not a good choice for improving the search algorithm. The other two values look more promising. Ref.~\cite{Nahimovs:2019} suggests to use $a=m-\sqrt{m}$ for large number of marked vertices. However, we can observe that the solid blue curves for $a=m$ and dashed yellow curves for $a=m-\sqrt{m}$ get close to each other as $m$ increases. Therefore, both values have the same behavior for large $m$.  
For small values of $m$, the dashed yellow curves $a=m-\sqrt{m}$ leads to smaller success probability and higher running time. }}
Overall $a=m$ behaves better and it can be a suitable choice for the self-loop weight in order to improve the search algorithm.
In the next section, we will analyse the behavior of the self-loop weight $l = m\cdot d/N$ when searching for an arbitrary set of marked vertices.

%
%
\subsection{Analysis for the self-loop weight $l=m\cdot d/N$}

\begin{figure}[!htb]
\centering
\subcaptionbox{}{\includegraphics[width=0.48\textwidth]{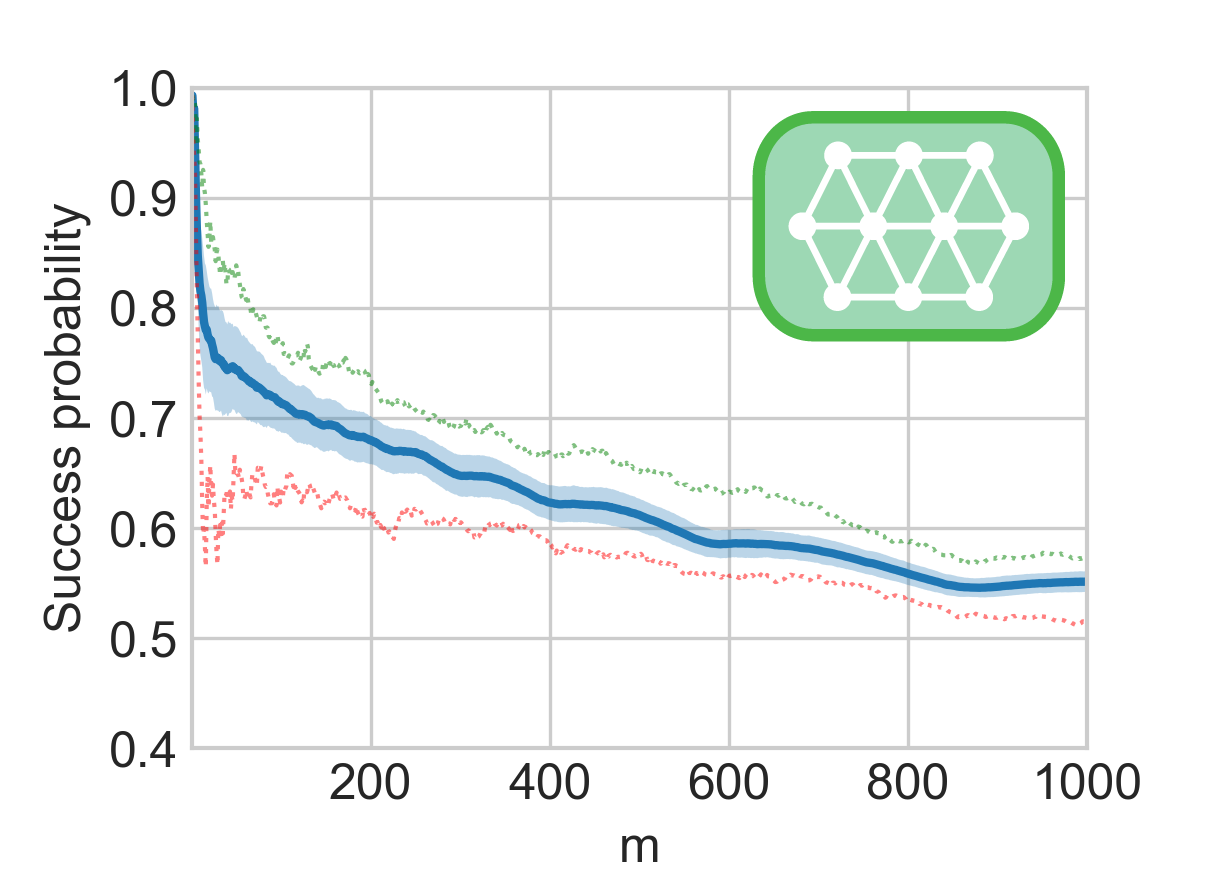}}
\hfill
\subcaptionbox{}{\includegraphics[width=0.48\textwidth]{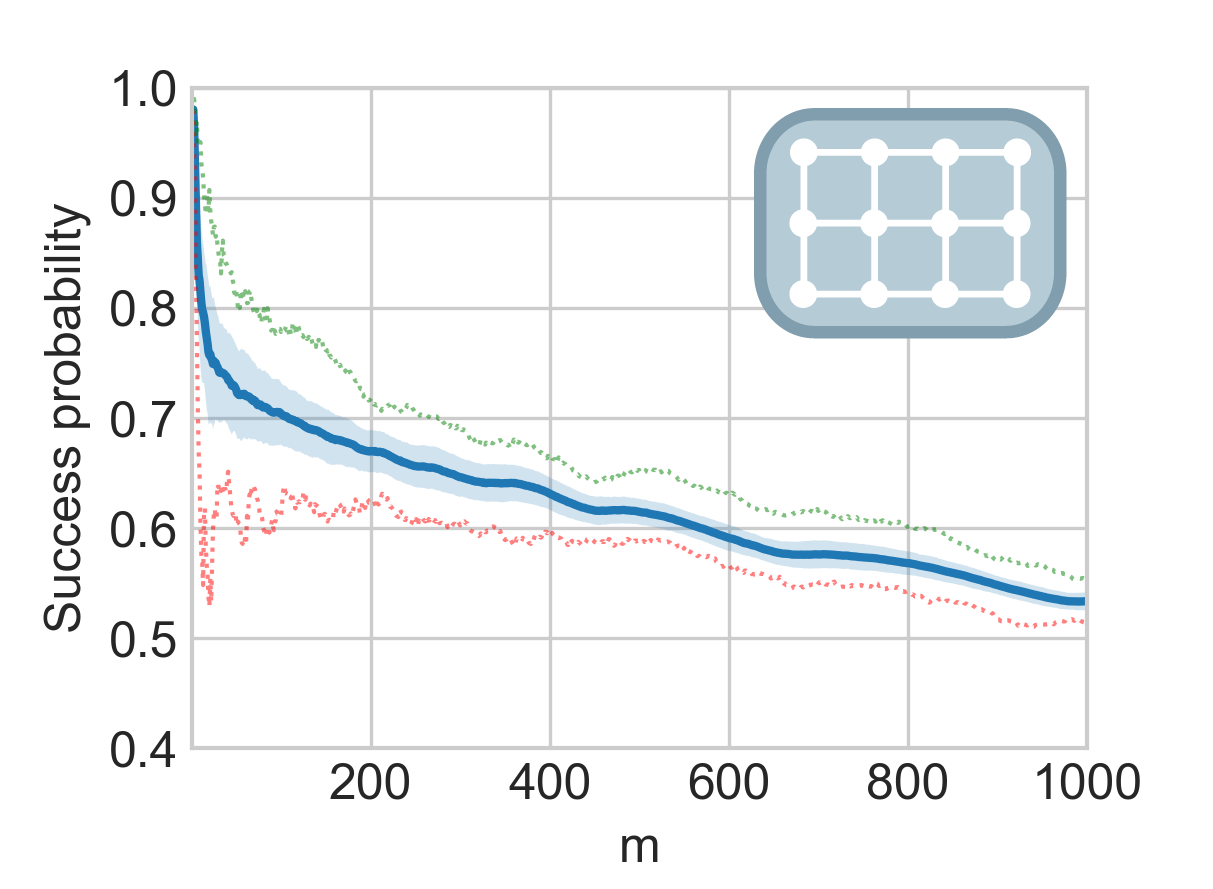}}
\hfill
\subcaptionbox{}{\includegraphics[width=0.48\textwidth]{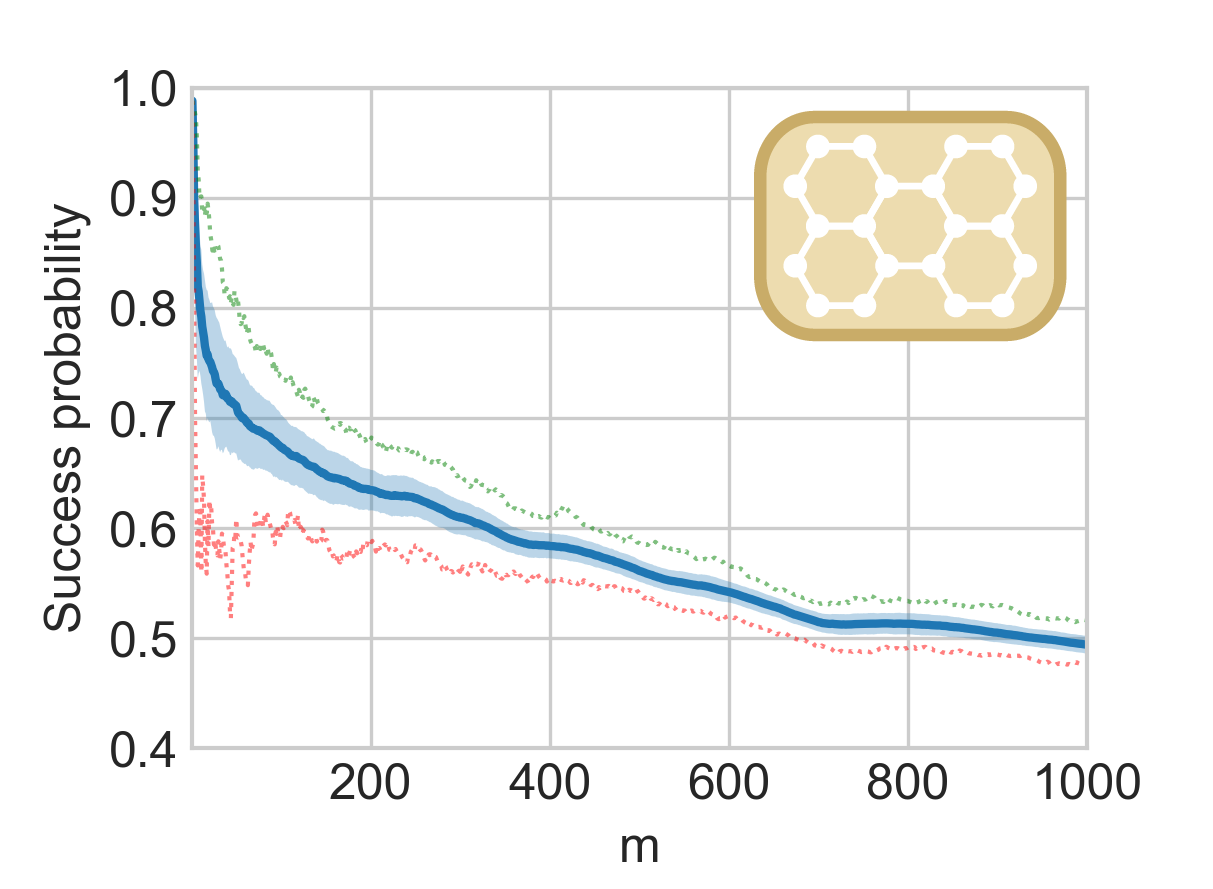}}
\caption{The average maximum success probability over 100 runs for the (a) triangular, (b) rectangular and (c) honeycomb grids. The self-loop weight is $l= m\cdot d/N$. The grids size is $N=100\times100$. The light blue area depicts the confidence band. The dotted light green and red curves are the maximum and minimum values, respectively, over the 100 runs.  The sets of marked vertices with sizes ranging from 1 to 1000 are randomly generated.}
\label{fig:prob}
\end{figure}

The following plots show the average over 100 executions of the search algorithm, considering randomly generated sets of marked vertices for $m \in [1,1000]$ for a grid of size $100\times 100$. Figure~\ref{fig:prob} depicts the average maximum success probability as a function of the number of marked vertices $m$.     
For all three types of grid the average success probability is higher and very close to 1 for $m=1$ and it is inversely proportional to the number of marked vertices (as noticed before in~\cite{Carvalho:2020} for the rectangular grid). For the shown range of marked vertices, the success probability stays above 0.45. The standard deviation is very small, which is shown by the confidence interval depicted by the light blue area. 

It is worth to mention that in our experiments sets of marked vertices were generated at random and we didn't prevent them to form an exceptional configuration\footnote{It is known, that lackadaisical quantum walks have exceptional configurations~\cite{Nahimovs:2019} of marked vertices. In such cases, the probability of finding a marked vertex stays close to the initial probability during the evolution.}.
This, however, did not affect the results. This can be seen from the values of the minimum probability (dotted light red curves) which are far from the initial probability. Note, that it is still possible that some smaller subsets of marked vertices have formed exceptional configurations.

\begin{figure}[!htb]
\centering
\subcaptionbox{}{\includegraphics[width=0.48\textwidth]{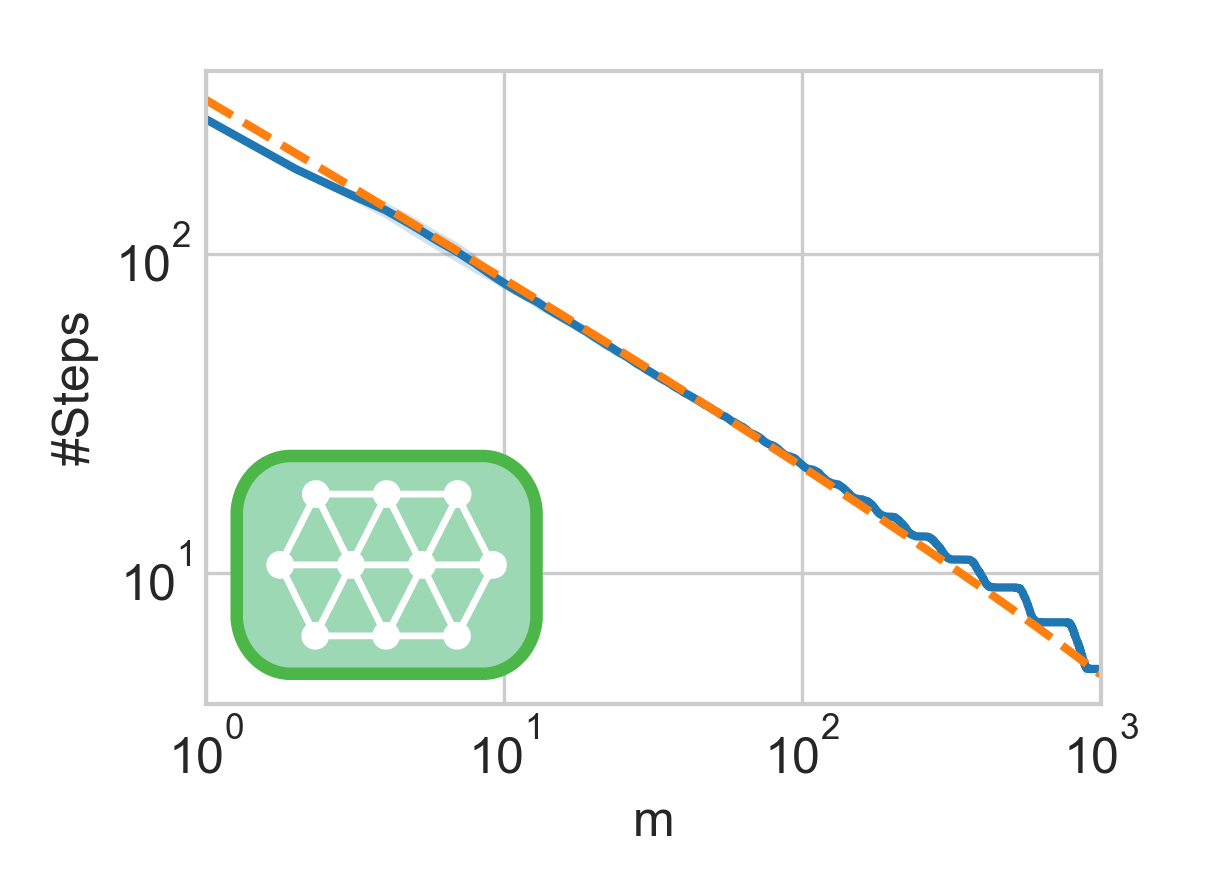}}
\hfill
\subcaptionbox{}{\includegraphics[width=0.48\textwidth]{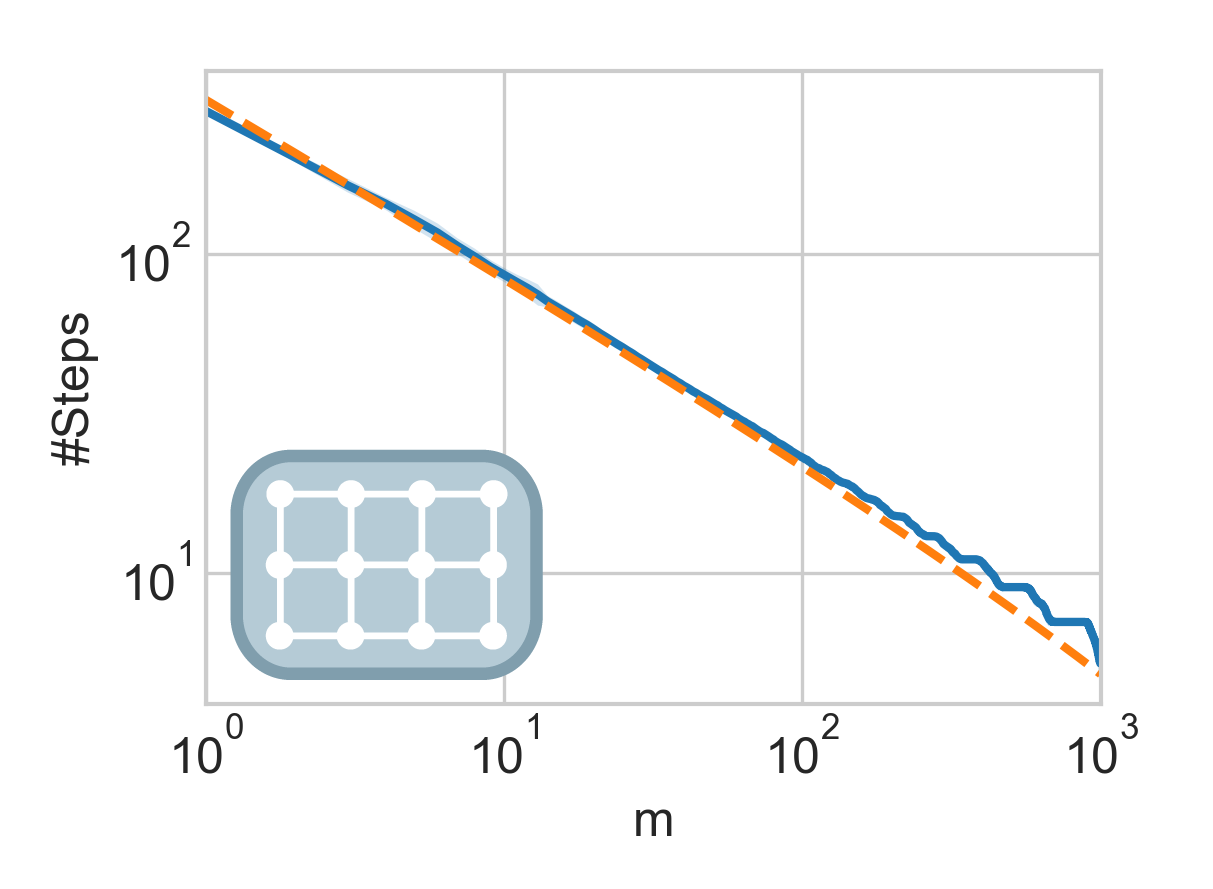}}
\hfill
\subcaptionbox{}{\includegraphics[width=0.48\textwidth]{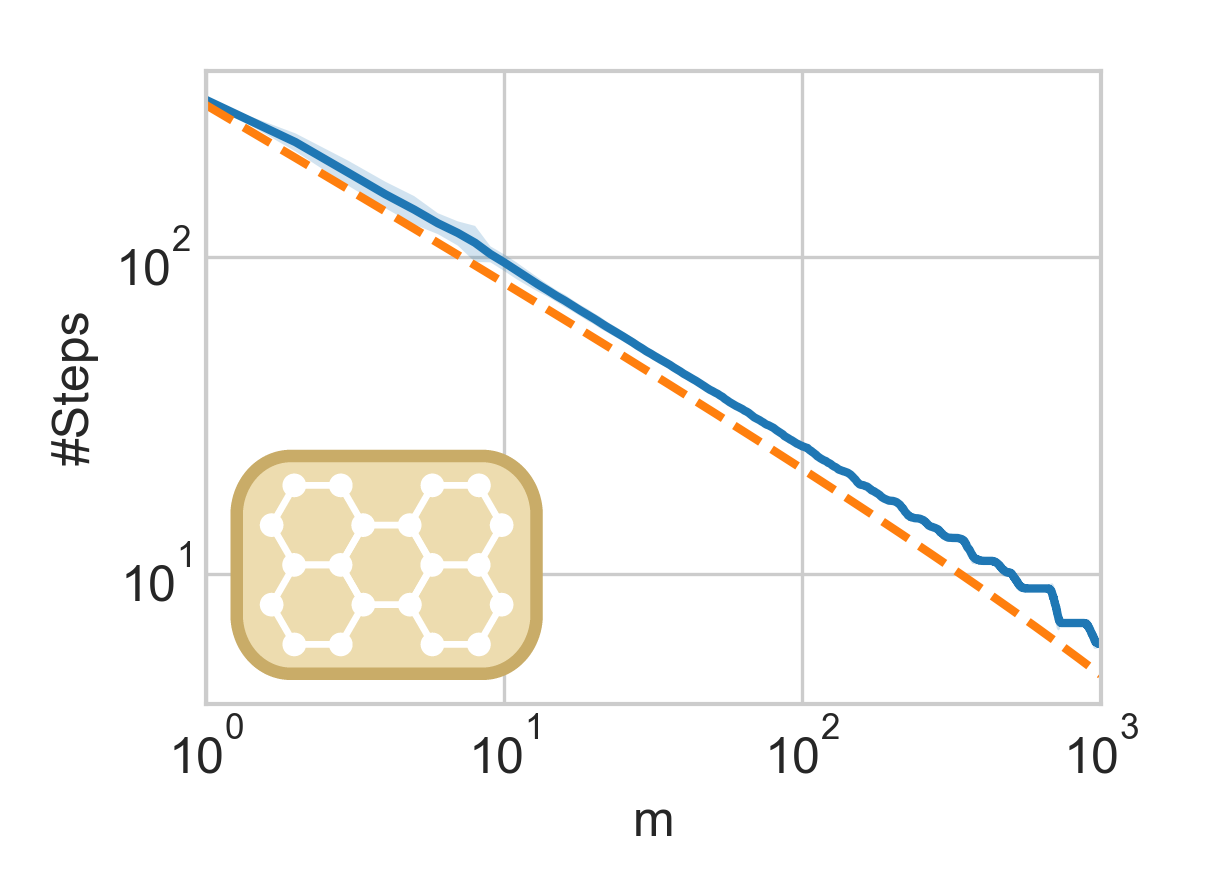}}
\caption{The average number of steps (solid blue curves) over 100 cases for the (a) triangular, (b) rectangular and (c) honeycomb grids in log-log scale. The self-loop weight is $l= m\cdot d/N$. The grids size is $N=100\times100$. The light blue area depicts the confidence band. The dashed yellow curve is $\sqrt{\frac{N}{m}\log \frac{N}{m}}$. The sets of marked vertices with sizes ranging from 1 to 1000 are randomly generated.}
\label{fig:steps}
\end{figure}

The average number of steps (the time to reach the first peak of the success probability) required by the algorithm is shown by Figs.~\ref{fig:steps}a, \ref{fig:steps}b and \ref{fig:steps}c for the rectangular, triangular and honeycomb grids, respectively. The plots are in log-log scale. The dashed yellow curve represents the function $\sqrt{\frac{N}{m}\log{\frac{N}{m}}}$. We can see that the curves for the number of steps and the dashed yellow curve have almost the same inclination. For the triangular and rectangular grids, the dashed yellow curve is almost a perfect fit. The same behavior can be seen when we scale the size of the grids. This means that the number of steps required by the search algorithm is $O\left(\sqrt{\frac{N}{m}\log{\frac{N}{m}}}\right)$. 

\begin{figure}[!htb]
\centering
\includegraphics[width=.6\textwidth]{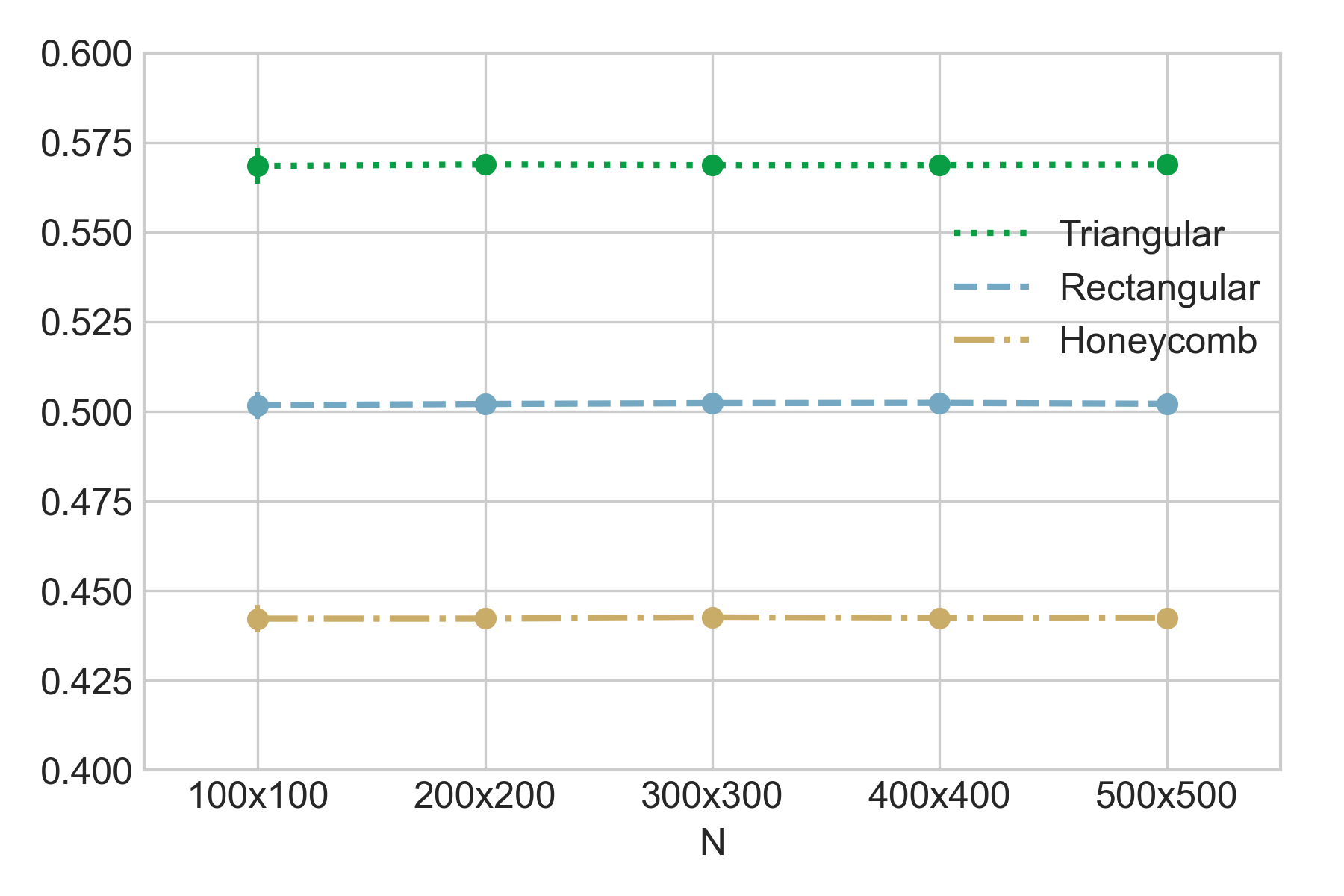}
\caption{The average maximum success probability  over 100 cases for different sizes of the triangular, rectangular and honeycomb grids. The self-loop weight is $l= m\cdot d/N$. The sets of marked vertices with size $m = N/5$ are randomly generated. }
\label{fig:prob2}
\end{figure}

Let us see what happens when we increase the number of marked vertices even further. Consider $20\%$ of vertices marked, that is,  $m=N/5$. Figure~\ref{fig:prob2} shows the average maximum success probability over 100 cases for the 3 types of grid with different sizes. We can see that the values stays above 0.4 for all grids. The honeycomb grid is the one with lower values, followed by the rectangular and then the triangular grid. The standard deviation is very small and it can be barely seen by the error bars.
In conclusion, since we know that the success probability is above some threshold, we can search with $O(1)$ probability in $O\left(\sqrt{\frac{N}{m}\log{\frac{N}{m}}}\right)$ time.


\section{Conclusions}\label{sec:conclusions}

We have studied the search for multiple marked vertices by lackadaisical quantum walks on triangular, rectangular and honeycomb 2D grids. Previous to our work, a few researchers \cite{Saha:2018,Giri:2020,Nahimovs:2019,Carvalho:2020} have studied the case of having multiple marked vertices. However, the previous research was focused on the rectangular grid and covered only a small range and specific sets of marked vertices. In this paper, we presented numerical results for different types of grids and a broad range and arbitrary sets of marked vertices.
We have demonstrated that by using lackadaisical quantum walks with self-loop weight of $l=m\cdot d/N$, we can find a marked vertices on triangular, rectangular and honeycomb 2D grids with $O(1)$ probability in $O\left(\sqrt{\frac{N}{m}\log\frac{N}{m}}\right)$ steps. The presented results are based on numerical simulations and it would be nice to obtain analytical results using the new method developed in~\cite{Bezerra:2021}.


\section*{Acknowledgements}
This work was supported by ERDF project number 1.1.1.2/VIAA/1/16/002 and by the QuantERA ERA-NET Cofund in Quantum Technologies implemented within the European Union's Horizon 2020 Programme (QuantAlgo project).


\bibliography{Paper}


\end{document}